\documentclass[aps,prl,amsmath,amssymb,10pt,twocolumn,superscriptaddress,notitlepage]{revtex4-1}
\usepackage{graphicx}
\usepackage{dcolumn}
\usepackage{bm}
\usepackage{amssymb}
\usepackage{braket}

\begin{document}

\title{Eliminating light shifts in single-atom optical traps}

\author{Nicholas R. Hutzler}
\email{hutzler@physics.harvard.edu}
\author{Lee R. Liu}
\author{Yichao Yu}
\author{Kang-Kuen Ni}
\email{ni@chemistry.harvard.edu}
\affiliation{Department of Chemistry and Chemical Biology, Harvard University, Cambridge, Massachusetts, 02138, USA}
\affiliation{Department of Physics, Harvard University, Cambridge, Massachusetts, 02138, USA}
\affiliation{Harvard-MIT Center for Ultracold Atoms, Cambridge, Massachusetts, 02138, USA}

\date{\today}

\begin{abstract}
Microscopically controlled neutral atoms in optical tweezers and lattices have led to exciting advances in the study of quantum information and quantum many-body systems.  The light shifts of atomic levels from the trapping potential in these systems can result in detrimental effects such as fluctuating dipole force heating, inhomogeneous detunings, and inhibition of laser cooling, which limits the atomic species that can be manipulated. In particular, these light shifts can be large enough to prevent loading into optical tweezers  directly from a magneto-optical trap.  We present a general solution to these limitations by loading, cooling, and imaging single atoms with temporally alternating beams.  Because this technique does not depend on any specific spectral properties,  we expect it to enable the optical tweezer method to control nearly any atomic or molecular species that can be laser cooled and optically trapped. Furthermore, we present an analysis of the role of heating and required cooling for single atom tweezer loading.
\end{abstract}
\maketitle

Interacting neutral atoms with quantum controls are a powerful platform for studies of quantum information and quantum many-body physics.  Systems of individually trapped atoms\,\cite{Schlosser2001} offer single particle control and detection with the flexibility to configure geometry and interactions in real time. This versatility has already allowed many proof-of-principle demonstrations, which include quantum logic gates\,\cite{Wilk2010,Isenhower2010,Muller2014Rydberg,Kaufman2015}, single atom switches of photons\,\cite{Tiecke2014}, and quantum simulators of spin systems\,\cite{Labuhn2015,Muller2014Rydberg}.  Scaling up the complexities of such a system by increasing the number of particles or species trapped offers exciting new directions. For example, dipolar atoms and polar molecules offer long-range, tunable, anisotropic interactions.   Molecules also possess many  internal degrees of freedom that provides additional handles for quantum control.
Arrays of individually controlled, ultracold dipolar atoms and polar molecules would allow explorations of new strongly correlated systems and exotic quantum phases\,\cite{Baranov2012}. 

One platform for realizing these applications is to confine single atoms in tight optical dipole ``tweezer" traps, where the size of the trap is of order the wavelength\,\cite{Schlosser2001}.   Since the polarizabilities of the ground and excited states are not perfectly matched, the atomic transitions will be shifted relative to their value in free space by a light shift\,\cite{Grimm2000}.  This gives rise to a number of undesirable effects when scattering near-resonant photons, such as fluctuating dipole force heating\,\cite{Dalibard1985,Alt2004}, where the atom sees  jumps in the gradient of the trapping potential as it cycles between the ground and excited state, inhibition of cooling due to the breakdown of hyperfine coupling\,\cite{Haller2015,Neuzner2015}, and spatially varying detuning and scattering rate.  Because cooling is required for loading and imaging, these effects can interfere with successful operation of the tweezer. Therefore, the successful loading of a wide variety of atomic species, each with an associated level structure, is made challenging by the effects of light shifts.

In this manuscript we investigate and clarify the roles of light shifts in loading, cooling, and imaging of atoms in optical tweezers, and demonstrate a general technique to eliminate them.  The technique utilizes fast (1 $-$ 3 MHz) modulation of the tweezer and resonant light to achieve not only single atom imaging\, \cite{Walker2012,Shih2013}, but also robust single atom loading for both cesium (Cs) and sodium (Na), the latter of which suffers from significant light shifts that would otherwise inhibit tweezer loading from a magneto-optical trap (MOT). We expect this technique to enable single atom and molecule loading of tweezers of essentially any species that can be laser-cooled and optically trapped.  Furthermore, this technique could  be applied to lattice imaging of individual atoms under a quantum gas microscope for atomic species where light shifts would prevent efficient photon scattering or inhibit cooling. 

\begin{figure}[t]
	\centering
		\includegraphics[width=0.47\textwidth]{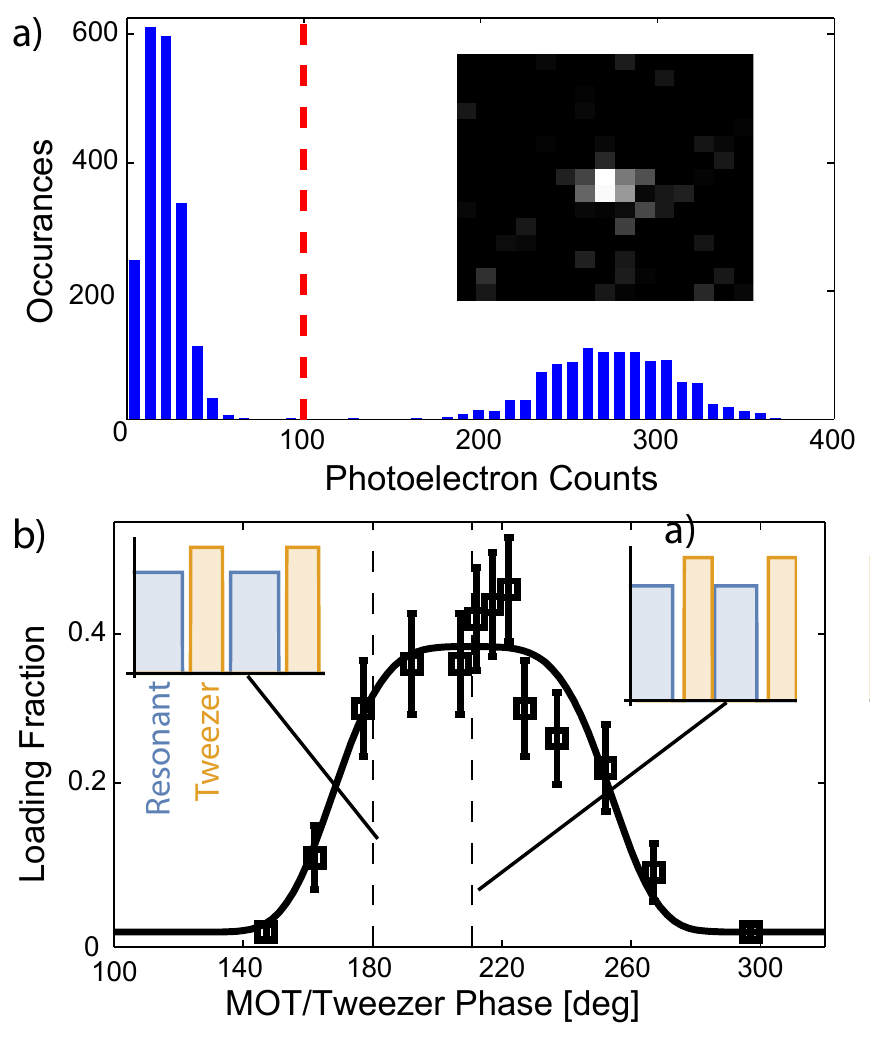}
	\caption{Single Na atom  in a tweezer. (a) A histogram of photoelectron counts in a 5x5 region around the atom for repeated loading cycles. The cutoff (red dashed line) distinguishes between 0 and 1 atom, which are the only outcomes for this loading mechanism\,\cite{Schlosser2001} (Inset) An image of a single Na atom in the tweezer. (b) Single Na atom loading performance for varying  relative phase between MOT and tweezer modulated beams.  When the resonant (MOT) and tweezer light overlap, light shifts prevent loading and imaging.  The data shown here is for 3 MHz modulation with 50\% duty cycle for the resonant light and 30\% for the tweezer light intensity, respectively.  The curve is to guide the eye. (Inset) Timing sequences of resonant and tweezer light at phase delays of 180$^{\circ}$ and 211$^{\circ}$ (optimum, corresponding to $\sim$ 30 ns).}
	\label{fig:SingleNaAtom}
\end{figure}

We investigate light shift effects for Na and Cs atoms in optical tweezers in an apparatus that follows the general approach of refs.\,\cite{Schlosser2001,Kaufman2012,Thompson2013}.  A collimated, red-detuned laser beam is incident on a 0.55 NA objective that creates a diffraction-limited, sub-micron tweezer.  The wavelength ranges used are $895-980$ nm for Cs, and $700$ nm for Na.  The focus of the objective is in the center of a MOT, which provides a local high density cloud of cold atoms for loading into the tweezer.  Atoms crossing the tweezer in the presence of cooling from the MOT beams may be loaded into the tweezer.  A dichroic mirror separates the tweezer light from fluorescence of the trapped atom, which is then focused onto a camera for imaging.  After loading the atom into the tweezer, the MOT cloud is allowed to disperse so that the atom can be imaged with a low background.  Single atoms are identified by imposing a threshold of photon counts (Figure  \ref{fig:SingleNaAtom}).  The same beams are used for the MOT, cooling, and imaging, and will generally be referred to as the ``resonant beams''.

\begin{figure}[t]
	\centering
		\includegraphics[width=0.47\textwidth]{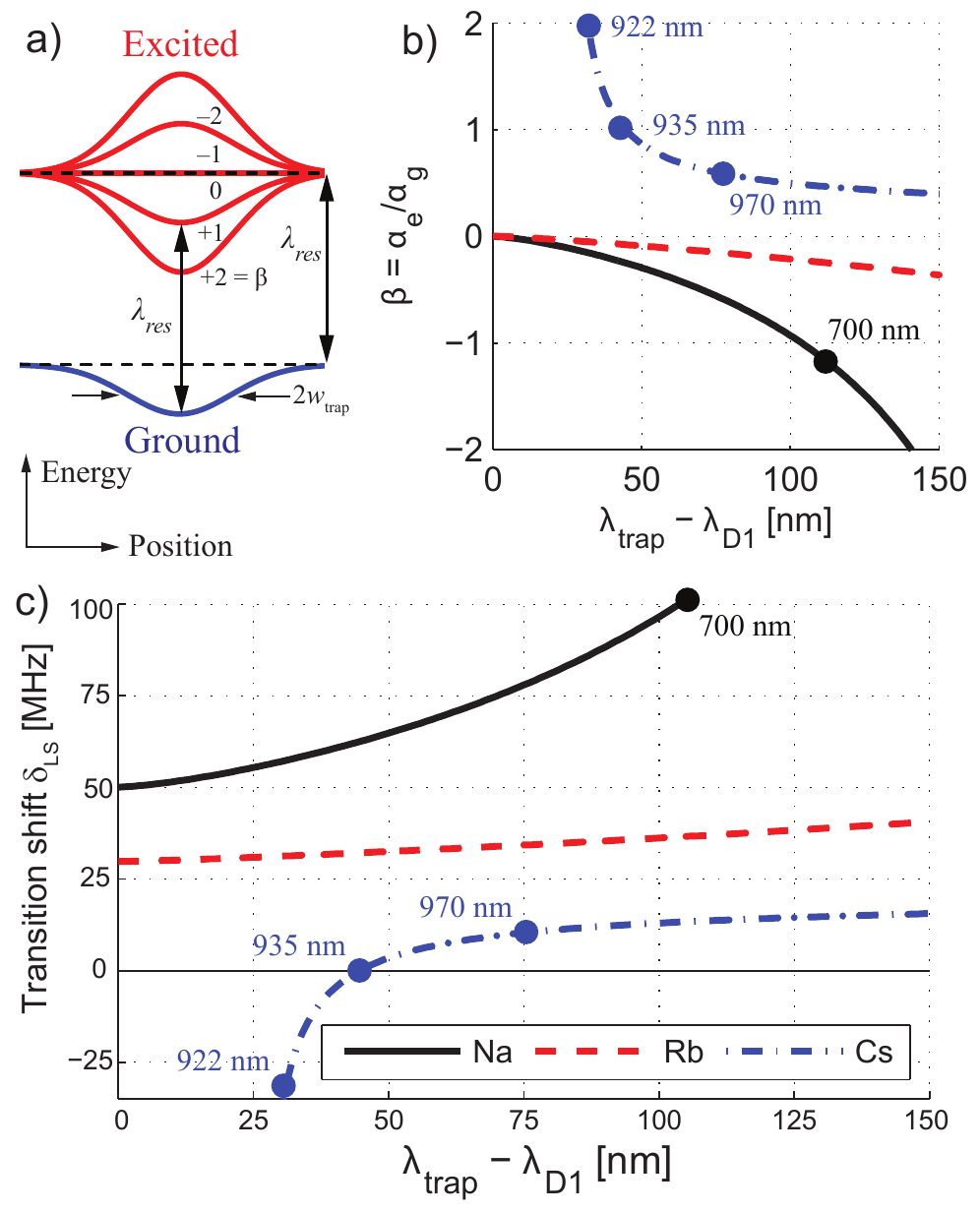}
	\caption{Light shifts in optical dipole traps. (a) An illustration of light shifts in a trap with waist $w_{trap}$ for ground (blue) and different excited state potentials (red) in terms of the excited/ground polarizability ratio $\beta=\alpha_e/\alpha_g$.  For $\beta<1$ $(\beta>1)$ the atom will see resonant light of wavelength $\lambda_{res}$ become red-shifted (blue-shifted) upon entering the trap. (b) $\beta$ for Na and Cs (Rb is also plotted for comparison).  We plot only the polarizability for the $m_j=3/2$ manifold\,\cite{Arora2007}. The wavelengths are plotted in reference to the $D1$ transitions for Na, Rb, and Cs which are 590 nm, 795 nm, and 895 nm respectively.
(c) A comparison of light shifts on the cycling transition for Na, Rb, and Cs atoms, following\,\cite{Arora2007}.  The trap depth is set to be $10\;T_{Doppler}$ for the ground state of each atom.  Transition light shift $\delta_{ls}$ is defined as the change in the transition frequency relative to the case in the absence of the trap; a positive shift means that the energy splitting between the ground and excited states increases (e.g. $\beta=-1$).    The light shifts for Na are large enough that hyperfine breakdown has already set in.   }
	\label{fig:ExcitedStateShifts}
\end{figure}

Due to the many electronic states in atoms, the polarizability of a given excited state, $\alpha_e$, can be either positive or negative independent of the ground state polarizability, $\alpha_g$\,\cite{Arora2007} (Figure \ref{fig:ExcitedStateShifts}a).  We define the wavelength-dependent ratio of polarizabilities as $\beta\equiv \alpha_e/\alpha_g$. In the special case when $\beta= 1$, a ``magic" wavelength\,\cite{McKeever2003,Katori1999}, the tweezer shifts the ground and excited state by equal amounts, and the atom experiences no light shifts. In Figure \ref{fig:ExcitedStateShifts}, we calculate light shifts for Cs and Na (and Rb for comparison) in the presence of a red-detuned tweezer of depth $10\;T_{dopp}\sim 1-3$ mK, where $T_{dopp}$ is the Doppler temperature, for a range of trapping wavelengths.  For Cs atoms in the range of $\sim 930-970$ nm, the light shifts are small, and are near zero ($\beta=1$) at 935 nm. For Na atoms over a large range of experimentally convenient wavelengths (630 nm $\sim$ 1064 nm), $\beta < 0$. 
Combined with the higher Doppler temperature of Na, this results in a large light shift that reduces the photon scattering rate and prevents the cooling that is required to capture the atom. Furthermore, the light shift is comparable to the excited state hyperfine splitting of $\approx 60$ MHz and  inhibits sub-Doppler cooling due to the breakdown of hyperfine coupling\, \cite{Haller2015,Neuzner2015}. Finally, attempting to load the atom from a MOT, where the excited-state fraction is typically $\sim 25\%$, an anti-trapped excited state will reduce the average trap depth, therefore requiring higher intensity and resulting in even larger light shifts and fluctuating dipole forces.

To circumvent issues related to loading, heating, and imaging that result from light shifts, we alternate the trapping and cooling light such that they are never on at the same time.  Specifically, we modulate the intensities of the tweezer and resonant light as square waves with frequencies between 1 and 3 MHz.  The fast modulation technique works well as long as the trap modulation frequency $f_{mod}$ is much greater than twice the trap frequencies, so the atom does not suffer from parametric heating\,\cite{Savard1997}, yet still experiences a time-averaged trap given by the average intensity.  In addition we require $f_{mod} \lesssim \gamma/2\pi$, where $\gamma$ is the natural linewidth, so that the atom will have enough time to decay into the ground state before the trapping light is switched back on.  A similar technique has been used in the past for light shift-free imaging of optically trapped atoms
\cite{Walker2012,Shih2013}.

The modulation is realized by using the first order diffracted beam from an acoustic-optical modulator driven by an 80 MHz sine wave mixed with the modulating square wave.  The resonant beams have 50\% duty cycle \footnote{We find that the resonant light can be modulated at all times and still yield a dense MOT with temperature $\lesssim 2T_{dopp}$, and that polarization gradient cooling with modulated beams yields temperatures similar to those achieved with unmodulated (CW) beams.  The lifetime of the single atom in the tweezer is $\approx 5$ seconds for both modulated and CW tweezers.}, and the tweezer has 30-40\% duty cycle to minimize overlap with the resonant light.  With this technique, single atoms were successfully loaded into a tweezer from a MOT or an optical molasses ($T\approx 10-30$ $\mu$K).  An image of a single Na atom and a histogram of photon counts from repeated loading attempts using the modulation technique is shown in Figure \ref{fig:SingleNaAtom}(a). We note that, in the absence of the modulation technique, we were not able to observe loading of a single Na atom from a MOT or molasses into a diffraction-limited tweezer\,\footnote{We were able to load into larger tweezers with waist $> 1$ micron with an unmodulated tweezer and MOT beams, though the loading was only a few percent efficient.}  after varying a wide range of parameters including tweezer depth, wavelength, MOT cooling power, repump power, detuning, and magnetic field gradient.

To illustrate the robustness of fast modulation and the detrimental effects of light shifts, we vary the relative phase of the resonant light and tweezer modulation and measure the probability of loading an atom in Figure \ref{fig:SingleNaAtom}(b).  When the tweezer and resonant light are not on at the same time, the atoms see no light shift but are still Doppler cooled, and we can reliably load the tweezer.  On the other hand, as the tweezer and resonant light begin to overlap in time, the light shifts inhibit photon scattering and the loading suffers.  We find that the center of the loading curve is not when the resonant light and tweezer are exactly out of phase (180$^\circ$), but with the tweezer trails resonant light turn-off by $\sim$ 30 ns due to the time that the atoms spend in the excited state (see inset of Figure \ref{fig:SingleNaAtom}b).

\begin{figure}[t]
	\centering
		\includegraphics[width=0.47\textwidth]{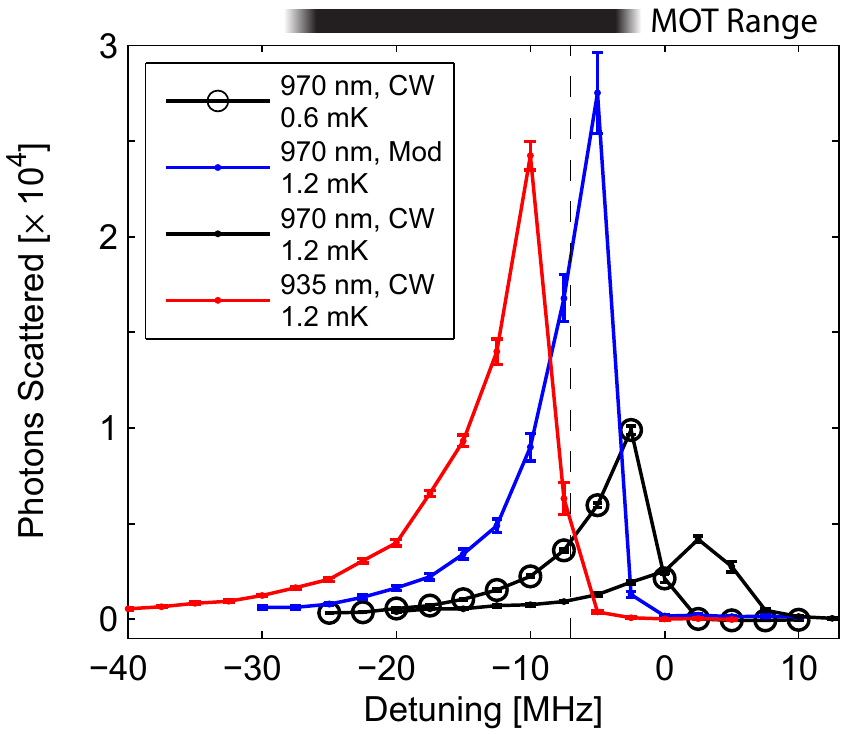}
	\caption{Scattered photons (assuming 4\% detection efficiency) vs. imaging detuning for single Cs atoms, with various combinations of modulated (Mod) vs. unmodulated (CW), 0.6 mK and 1.2 mK tweezer depths, and 970 nm or 935 nm tweezer wavelength.  The modulated data means that there is effectively no light shift.  The MOT detuning (for single atom loading experiments) is indicated by the vertical dashed line at -7 MHz.  0 MHz corresponds to the free-space atomic resonance. The qualitative line shape is explained in the main text.} 
	\label{fig:PhotonsVsDetuningCurve}
\end{figure}

To understand the roles of light shifts in loading and imaging, we can study the number of photons that an atom can scatter in the tweezer versus detuning.  Figure \ref{fig:PhotonsVsDetuningCurve} shows photons scattered versus detuning $\delta$ (relative to the atom in free space) for a single Cs atom in a tweezer with an  imaging duration of 50 ms and an intensity of $0.3$ mW/cm$^2\approx 0.1 I_{sat}$.    While illuminated with near-resonant light, the atom scatters photons  at a rate that depends on the detuning from the atomic resonance\,\cite{Metcalf1999}, and experiences recoil heating due to spontaneous emission.  Applying the modulation technique to imaging single atoms gives a reference line shape that is free of light shifts.  For ${\delta \gtrsim -\gamma/2}$, no effective cooling is present and therefore only a small number of photons can be scattered before the atom is heated out of the tweezer\,\footnote{With no cooling, the atom would scatter typically on the order of 100 photons before being heated out of the $\approx$ 1 mK tweezer}. 
However, if the near-resonant light is red-detuned on the order of $\delta \lesssim-\gamma/2$, then Doppler and sub-Doppler cooling can keep the atom cold while it scatters photons.  We find the equilibrium temperature $T_{eq}$ is typically around $1/4-1/3$ of $T_{dopp}$ (with either CW or modulated beams), which is well below the $U_0\approx$ 1 mK tweezer depths used here\,\footnote{Note that even when $T_{eq} < U_0$, the atom will eventually escape the tweezer; there is a finite fraction of the Boltzmann distribution with temperature $T_{eq}$ that is above $U_0$, and as the atom scatters photons and samples the distribution it will eventually reach an energy that is above $U_0$ and is therefore no longer trapped.  Numerical estimates show that $U_0\approx 10\times T_{dopp}$ is generally sufficient to scatter the 1,000-10,000 photons needed for high-fidelity imaging with a single shot.}. As the detuning becomes more red, the number of photons scattered is decreased due to the finite imaging time.  A numerical model of the line shape is given in the appendix.

To quantify and illuminate the roles of different heating and cooling effects due to light shifts, we further combine measurements that introduce a controlled amount of light shifts to the Cs atom by tuning the tweezer wavelengths and  depths without modulation. 
When a light shift $\delta_{ls}$  is present, the atomic resonance shifts accordingly. In Fig. \ref{fig:PhotonsVsDetuningCurve}, the peaks of the 970 nm CW tweezers for two depths track the $\delta_{ls}$ shift while the scattering line shapes qualitatively retain the same asymmetry - cooling on the red side (left) of the peak and heating on the blue side (right) of the peak.   
Furthermore, the peak number of photon scatters reduces as the light shifts increase due to fluctuating dipole force heating and inhomogeneous detuning, that is, the fact that the atom will see a range of detunings as it samples different trap depths.  For $\beta = 1$ (magic wavelength at 935 nm), the peak photon number is similar to the no light shift case. The residual shift of the 935 nm curve is likely due the fact that the magic wavelength is not for all hyperfine levels.

The scattering line shapes in Fig. \ref{fig:PhotonsVsDetuningCurve} not only provide information about single atom imaging, but also  crucially connects to single atom loading, since the conservative tweezer potential requires cooling in order to trap an atom. A numerical estimate suggests that of the order 100 photons are required to cool the atom into the trap.  During single atom loading, the cooling provided by the resonant light has a detuning that is constrained relative to the free space value ( $-$7 MHz for Cs in our experiment) since the MOT has a constant detuning.  This detuning can be adjusted to match the light shift, but is limited to a finite range for reliable MOT loading (shaded bar in Fig. \ref{fig:PhotonsVsDetuningCurve}). 
The regimes where $\beta>1$ and $\beta<1$ present different challenges to atom loading. For $\beta>1$, the atom will see the resonant beams become shifted to the blue upon entering the tweezer ($\delta_{ls}<0$).  If $\beta$ is large enough such that $|\delta_{ls}|\gtrsim|\delta_{MOT}|$, this will result in significant Doppler heating, and the atom cannot be efficiently loaded directly from a MOT.  We demonstrate this with Cs in a 922 nm tweezer, where $\beta\approx 2$; at this wavelength we were not able to load any single atoms using the conventional CW loading method, but achieved robust loading ($\sim$50\% success rate) with fast modulation due to the effective elimination of light shifts.

\begin{figure}[t]
	\centering
		\includegraphics[width=0.47\textwidth]{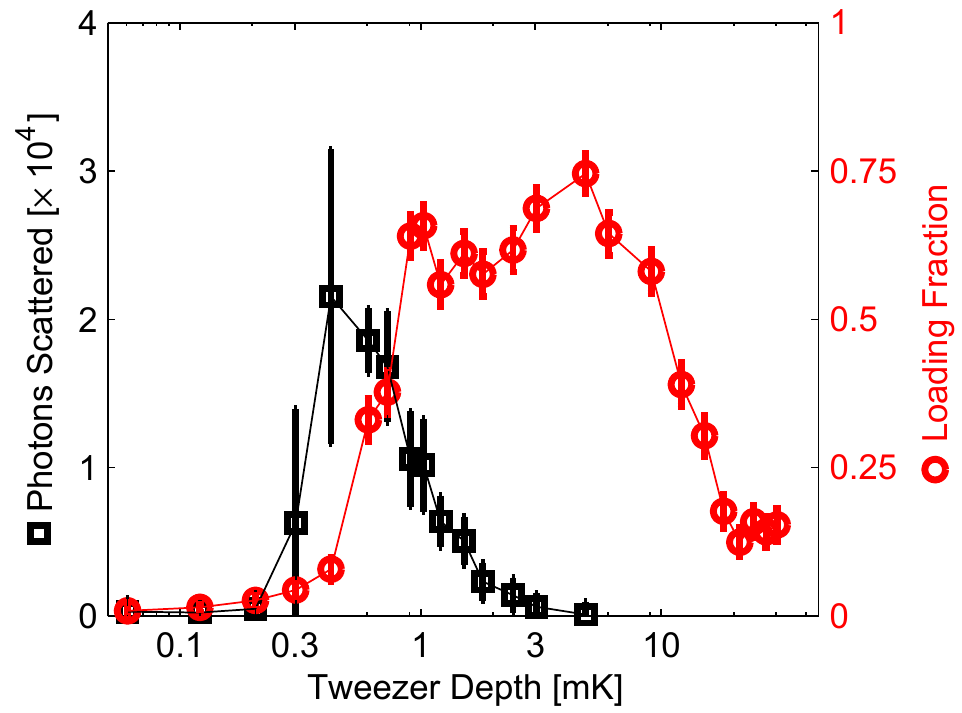}
	\caption{Photons scattered and loading probability for various CW tweezer depths. Both quantities require some minimum trap depth to allow sufficient photons to be scattered for detection.  The number of photons scattered decreases with increasing depth because light shifts reduce the scattering rate while the imaging time is kept fixed.  However, the loading fraction remains large as long as the scattering rate is large enough to cool the atom into the conservative trap.}
	\label{fig:PhotonsLoadingVsDepth}
\end{figure}

On the other hand, if $\beta < 1$, the atom will see the resonant light become shifted to the red in the tweezer ($\delta_{ls}>0$).  As long as this shift is not too large, Doppler cooling will continue and the atom can be loaded and imaged.  However, if the light shift is too large, the atom may not scatter enough photons to become deeply trapped.  Na atoms with a 700 nm tweezer ($\beta$ between $-1$ and $-2$ depending on hyperfine level)  falls into this category as discussed  prior to Fig. \ref{fig:SingleNaAtom}. Here, we demonstrate the breakdown of single Cs atom loading into a 970 nm tweezer ($\beta$ between $0$ and $0.5$ depending on hyperfine level)  as the trap depth (as well as the light shift) increases (Figure \ref{fig:PhotonsLoadingVsDepth}). We also measure how many photons can be scattered at  various corresponding trap depths. To eliminate variability in loading for the scattering rate measurement,  we load single atoms under a fixed trap depth ($\approx$ 1 mK) and ramp the  tweezer to various depths for imaging.  Imaging intensity and duration are kept fixed.  In Fig. \ref{fig:PhotonsLoadingVsDepth}, we  see that as the tweezer becomes deeper, the scattering rate is reduced due to the light shift that increases the effective detuning of the imaging light.  Similarly, resonant light becomes increasingly detuned during the loading phase as the atom is cooled into the tweezer and sees an increasing light shift.  For deep enough tweezers, the light shift increases so quickly that the scattering rate is turned off before the atom is effectively trapped.  Because fewer photons are needed to cool (of the order 100) compared to the number needed for high-fidelity images (of the order 1,000-10,000), the number of photons scattered falls more quickly than the loading rate as the trap depth is increased.

In conclusion, we present an experimental investigation the effects of light shifts in cooling, imaging, and trapping single atoms in optical tweezers, and demonstrate a general technique to eliminate light shifts. Our technique allows us to reliably load both single Na and Cs atoms in tweezer traps, which provide a promising avenue to produce single molecules. Our technique is versatile and can be applied to other interesting atomic and molecular species that can be optically trapped and cooled\,\cite{Barry2014}. This could provide novel sources of cold atoms and molecules for quantum information, quantum simulation, and to be interfaced with hybrid quantum systems.

We thank Adam Kaufman, Jeff Thompson,  and Mikhail Lukin for many helpful discussions, as well as Yu Liu and Jessie Zhang for experimental assistance.  N. R. H.  acknowledges support from Harvard Quantum Optics Center.  This work is supported by the NSF through the Harvard-MIT CUA, as well as the AFOSR Young Investigator Program, the Arnold and Mabel Beckman Foundation, the Alfred P. Sloan Foundation, and the William Milton Fund. 

\textbf{Appendix}

\appendix

\emph{Model for photons scattered vs. detuning}.  Consider an atom of mass $m$ in a 1D harmonic trap of depth $U_0\approx 1$ mK and temperature $T$.  Expose the atom to near-resonant light of wavelength $\lambda_{res}$ so that it begins to scatter photons at a rate $R_{scat}$ given by\cite{Metcalf1999}
\begin{equation}
R_{scat} = \frac{1}{2}\frac{s_0 \gamma}{1+s_0+(2\delta/\gamma)^2},
\end{equation}
where $\gamma$ is the natural width, $\delta$ is the detuning from resonance including light shifts, and $s_0=I/I_{sat}$ is the saturation parameter.  Let's consider the effects of Doppler heating/cooling, recoil heating, and polarization gradient cooling.

For $s_0 \ll 1$, the Doppler heating/cooling rate is given approximately by
\begin{equation}
\dot{E}_{dopp} = \langle \vec{F}_{OM}\cdot\vec{v} \rangle = \alpha \langle v^2\rangle = \alpha k_B T/m,
\end{equation}
where $\vec{F}_{OM} = -\alpha \vec{v}$ is the optical molasses force, and 
\begin{equation}
\alpha = \frac{8\hbar k^2 \delta s_0}{\gamma(1+s_0+(2\delta/\gamma)^2)^2}
\end{equation}
where $k = 2\pi/\lambda_{res}$.   We used the fact that $\langle v^2\rangle = k_B T/m$ in a 1D trap.  The recoil heating rate is given by
\begin{equation}
\dot{E}_{recoil} = 4\hbar\omega_{recoil}R_{scat},
\end{equation}
where $\omega_{recoil} = \hbar k^2/2m$.

\begin{figure}[htbp]
	\centering
		\includegraphics[width=0.45\textwidth]{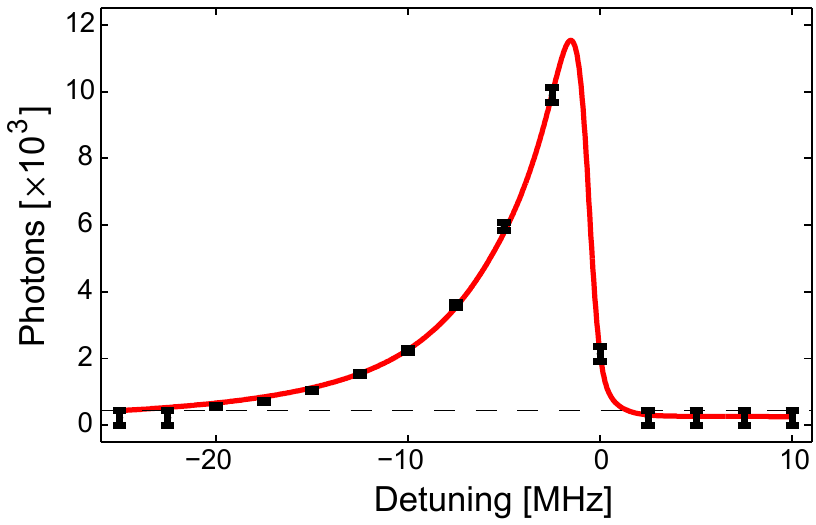}
	\caption{Fit of measured photons vs. detuning curve to the model.  We allow the overall height, light shift, and scaling factor on the heating/cooling rate to vary in the fit.  The data is for a single Cs atom in a 970 nm tweezer trap that is 0.6 mK deep, with initial temperature of 10 $\mu$K from polarization gradient cooling before imaging, which was measured independently by release/recapture and Raman sideband thermometry.  The horizontal dashed line indicates the detection limit due to background.}
	\label{fig:PhotonsVsDetuningFit}
\end{figure}

To model polarization gradient cooling (PGC) \cite{Dalibard1989}, we use
\begin{equation}
\dot{E}_{PGC} \propto T\hbar k^2 \frac{\delta\gamma}{5\gamma^2 + 4\delta^2},
\end{equation}
with a scaling factor chosen to reproduce the observed equilibrium imaging temperature of $\approx 40$ $\mu$K $<T_{dopp}$ for Cs.  Including PGC is important not only to understand the sub-Doppler temperature, but also the shape of the curve shown in Figure \ref{fig:PhotonsVsDetuningFit}.

The total heating/cooling rate of the atom $\dot{E}_{tot}$ is obtained by summing these contributions.  We can perform a simple estimate of the total number of photons scattered with the following routine, starting with some initial temperature $T_0$ and initial survival probability $P_0 = 1$:

\begin{enumerate}
\item Increase the temperature to $T_{i+1} = T_i + dt\times\dot{E}_{tot}/k_B$
\item Reduce the survival probability of the atom $P_i$ to $P_{i+1} = P_i\times(1-e^{-U_0/k_BT})$, the fraction of the Boltzmann distribution that is higher than the trap depth
\item Repeat until $P\ll 1$.
\end{enumerate}
If we use $dt = 1/R_{scat}$, then the total number of photons is given approximately by $\sum_i P_i$.  

This approach is very simple and ignores many of the complexities of the system, but captures the important features.  In particular, this model reproduces the overall shape of the photon vs. detuning data and helps build understanding of the loading and imaging mechanisms.

\bibliography{library}

\end{document}